\documentclass[12pt]{article}
\usepackage{epsfig,amssymb,amsmath,psfrag}

\def \bl  {\begin{align*}}
\def \el  {\end{align*}}

\def \be  {\begin{equation}}
\def \ee  {\end{equation}}
\def \ba  {\begin{eqnarray}}
\def \ea  {\end{eqnarray}}
\def \baa {\begin{eqnarray*}}
\def \eaa {\end{eqnarray*}}
\def \bb  {\begin {thebibliography} }
\def \eb  {\end{thebibliography}}
\def \lab #1 {\label{#1}}

\def \qqquad {\qquad\quad}

\renewcommand{\a}{\alpha}
\newcommand{\da}{{\dot{\alpha}}}
\newcommand{\db}{{\dot{\beta}}}
\renewcommand{\b}{\beta}

\def\XXint#1#2#3{{\setbox0=\hbox{$#1{#2#3}{\int}$}
     \vcenter{\hbox{$#2#3$}}\kern-.5\wd0}}

\def\l<{\langle}\def\r>{\rangle}

\renewcommand{\title}[1]{\vbox{\center\LARGE{#1}}\vspace{5mm}}
\renewcommand{\author}[1]{\vbox{\center#1}\vspace{5mm}}
\newcommand{\address}[1]{\vbox{\center\em#1}}
\newcommand{\email}[1]{\vbox{\center\tt#1}\vspace{5mm}}

\def \uu  {\bar}

\begin{document}

\begin{titlepage}
\begin{center}
\vspace{5mm}
\hfill {\tt HU-EP-09/06}\\
\hfill {\tt LAPTH-1308/09} \\
\vspace{20mm}

\title{\sc Yangian Symmetry of Scattering Amplitudes in $\mathcal{N}=4$ super Yang-Mills Theory}
\author{\large James Drummond${}^{1}$,  Johannes Henn${}^{2}$ and Jan Plef\/ka${}^{2}$}
\address{${}^{1}$LAPTH, Universit\'{e} de Savoie, CNRS,\\
B.P. 110,  F-74941 Annecy-le-Vieux Cedex, France\\[1cm]
${}^{2}$Institut f\"ur Physik, Humboldt-Universit\"at zu Berlin,\\
Newtonstra{\ss}e 15, D-12489 Berlin, Germany
}

\email{drummond@lapp.in2p3.fr,\{henn,plefka\}@physik.hu-berlin.de}

\end{center}

\abstract{
\noindent
Tree-level scattering amplitudes in $\mathcal{N}=4$ super Yang-Mills theory have
recently been shown to transform covariantly with respect to a `dual' superconformal symmetry algebra,
thus extending the conventional superconformal symmetry algebra $psu(2,2|4)$ of the theory.
In this paper we derive the action of the dual superconformal generators in on-shell
superspace and extend the dual generators suitably to leave scattering amplitudes invariant.
We then study the algebra of standard and dual symmetry generators and
show that the inclusion of the dual superconformal generators lifts the $psu(2,2|4)$
symmetry algebra to a Yangian. The non-local Yangian generators acting on amplitudes
turn out to be cyclically invariant due to special properties of $psu(2,2|4)$. The representation of the
Yangian generators takes the same form as in the case of local operators, suggesting that the
Yangian symmetry is an intrinsic property of planar  $\mathcal{N}=4$ super Yang-Mills,
at least at tree level.\vfill
}

\end{titlepage}

\section{Introduction}
The $\mathcal{N}=4$ supersymmetric Yang-Mills theory (SYM) \cite{N4SYM} is a remarkable model of mathematical
physics. To begin with it is the gauge theory with maximal supersymmetry and it is superconformally invariant
at the classical and quantum level with a coupling constant free of renormalisation.
In its planar limit all observables
of the $SU(N)$ model depend on a single tunable parameter, the 't Hooft coupling $\lambda$. Moreover the theory is dual to superstring theory on $AdS_5\times S^5$ in the sense of the
AdS/CFT correspondence \cite{AdS/CFT}.
In recent years we have seen formidable progress in our understanding of this most symmetric
AdS/CFT system  due to its hidden integrability  \cite{AdS/INT1,AdS/INT1b,AdS/INT2}.
{ The first hints of integrable structures in planar QCD appeared in the study of high-energy scattering processes \cite{Lipatov:1993yb,Faddeev:1994zg}.
In ${\mathcal N}=4$ SYM} integrability has been observed in the
study of anomalous scaling dimensions of gauge invariant local operators and in the spectral
problem of the classical and quantum $AdS_5\times S^5$ string.
On the gauge theory side the problem of finding the all-loop scaling dimensions
can be mapped to the eigenvalue problem of a long-range
integrable spin chain model, which in turn is governed by a nested set of
Bethe equations \cite{Betheeq}
(see \cite{intrevs} for reviews), for recent progress see \cite{GKV}. Here the spin chain Hamiltonian - alias the dilatation operator - acts on the
chain of fundamental fields - the `spins' - inside the trace of the local operator. Integrability
refers to the existence of an infinite set of charges, which are in involution, and contain the
Hamiltonian of the system. 
Integrability on the string side has been established at the level of the classical
theory, where again an infinite set of conserved charges may be constructed \cite{AdS/INT2,StringInt}.
A number of positive tests of the Bethe equations in the strong coupling limit
have been performed in various perturbative studies of the quantised $AdS_5\times S^5$ string,
see e.g.~\cite{SCTests}.
A natural question following these tremendous advances is, what implications integrability
has beyond the realm of two-point functions of local operators or, respectively, the excitation
spectrum of superstrings in an  $AdS_5\times S^5$ background.

Parallel and initially unrelated progress has been made in the study of on-shell scattering
amplitudes in $\mathcal{N}=4$ SYM. Maximally helicity violating (MHV) $n$-gluon
scattering amplitudes at tree-level have a very simple form when expressed in the spinor helicity
formalism \cite{PTBG}. As shown by Witten \cite{Witten:2003nn} at tree level
 there is an interesting perturbative duality to strings in twistor space.
This inspired the development of recursive techniques to construct
tree-level gauge theory amplitudes known as the BCFW recursion relations \cite{Britto:2004ap,Britto:2005fq}.
At the level of loop corrections Bern, Dixon and Smirnov (BDS)  \cite{Bern:2005iz}
conjectured an all-loop form of the MHV amplitudes
in $\mathcal{N}=4$ gauge theory, based on an iterative structure \cite{Anastasiou:2003kj} found at lower loop levels employing
on-shell techniques \cite{Bern:1994zx,Bern:1994cg} (see \cite{OSrev} for a recent review). The form of this conjecture was dictated by the tree-level and one-loop
structure and the
cusp anomalous dimension,  a quantity which is simultaneously the leading ultraviolet singularity of Wilson loops with light-like cusps \cite{Polyakov:1980ca,Korchemsky:1987wg,Korchemskaya:1992je} 
and the scaling dimension of a particular class of local operators in the high spin limit. Due to this latter property it may be obtained 
from the above mentioned Bethe equations. In fact the BDS conjecture is  
now known to fail at two loops 
and six points \cite{Alday:2007he,Drummond:2007bm,Bartels:2008ce,Bern:2008ap,Drummond:2008aq} but the deviations from it are constrained by a novel symmetry, dual conformal symmetry, which we will now discuss.

The dual string theory prescription for computing scattering amplitudes was
proposed in \cite{Alday:2007hr}.
There, the problem of computing certain scattering amplitudes at strong
coupling was mapped via a T-duality transformation to that of computing Wilson loops with light-like segments.
The strong coupling calculation of \cite{Alday:2007hr} is insensitive to the helicity structure of the
 amplitude being calculated. At weak coupling however, it was found that light-like Wilson loops are dual to MHV amplitudes, as was
demonstrated for $n=4,5,6$ gluons scattering up to two-loop order \cite{Drummond:2007aua,Brandhuber:2007yx,Drummond:2007cf}.
A direct implication of this Wilson
loop/MHV amplitude duality is the existence of a novel `{dual conformal}' symmetry
of the amplitudes, for which hints had appeared at the level of loop integrals
contributing to the amplitudes \cite{Drummond:2006rz,Bern:2006ew,Bern:2007ct}.
This symmetry has its interpretation as the ordinary conformal symmetry of the
Wilson loops \cite{Drummond:2007cf,Drummond:2007au}.
Indeed the two-loop, six-point calculations of \cite{Drummond:2007bm,Bern:2008ap,Drummond:2008aq} showed that the amplitude and the Wilson loop both
differ from the BDS conjecture by the same function of dual conformal invariants.
This symmetry extends naturally to dual superconformal symmetry when one considers writing all amplitudes (MHV and non-MHV) in on-shell superspace \cite{dhks5,dhks6}.
In particular, tree-level amplitudes are covariant under dual superconformal symmetry.
On the string side this symmetry can be seen to arise from the
combination of a bosonic and a novel fermionic T-duality transformation of the $AdS_5\times S^5$
superstring onto itself.
This transformation maps the dual superconformal symmetry of the original theory to the ordinary superconformal
symmetry of the dual model, thereby providing another indication for the dual superconformal symmetry of planar
scattering amplitudes \cite{Berkovits:2008ic,Beisert:2008iq}.

For tree-level amplitudes, dual superconformal symmetry was verified recursively \cite{Brandhuber:2008pf} by employing a supersymmetrised version \cite{ArkaniHamed:2008gz,Brandhuber:2008pf,Elvang:2008na} of the BCFW recursion
relations \cite{Britto:2004ap,Britto:2005fq}. In fact the supersymmetric recursion relations can be explicitly
solved to obtain a closed form for all tree-level amplitudes in $\mathcal{N}=4$ SYM \cite{Drummond:2008cr}.
The expression thus obtained is given by a sum over dual superconformally covariant quantities.

The natural question arising from these two symmetry algebras is, what mathematical
structure arises, when one commutes generators of the superconformal and dual superconformal
algebras with each other? One would indeed expect to generate an infinite-dimensional
symmetry algebra as a manifestation of the integrability
of the theory. We shall show in this paper that a Yangian symmetry of scattering amplitudes appears, at least
 at tree level.
We take our inspiration from the discussion of the behaviour of the infinite tower of charges
in the sigma model in references \cite{Berkovits:2008ic,Beisert:2008iq}\footnote{We thank Niklas Beisert for
important comments on this point, see also
 \cite{Beisert-talk}.}. There it
 was shown how the combination of bosonic and fermionic T-dualities maps the charges of the
 original model into the charges of the T-dual model. In this way dual superconformal symmetry
 can be identified with certain non-local charges which map to local ones.
Yangian symmetry has been observed in other instances of the AdS/CFT system,
at strong coupling in string theory \cite{AdS/INT2}, at weak coupling in the gauge theory \cite{Dolan:2003uh,Dolan:2004ps}, perturbatively in the spin-chain picture \cite{YangianChain},
as well as in the AdS/CFT S-matrix \cite{YangianSM}.
These occurrences have all been connected to the integrable spin chain picture and its
`world-sheet' S-matrix of magnon scattering along the chain. 
{ Traditionally Yangian symmetry is connected to integrable 2d field theories or spin chains,
for reviews see e.g.~\cite{Yangianrevs}.
However it is reasonable to expect that scattering amplitudes should exhibit such a symmetry, as suggested in \cite{Witten:2003nn}.}
Remarkably, we indeed find the emergence of a Yangian structure at the
level of the gauge theory scattering amplitudes  (i.e. the \emph{spacetime} S-matrix).
As we shall see we may again associate a `spin-chain' picture to the case of
scattering amplitudes, where the `sites' are identified with the on-shell external fields
in the colour ordered amplitude. This chain is by definition periodic and cyclically invariant.
The Yangian generators act non-locally along this chain and, as we shall show, are
also cyclic invariant when acting on a scattering amplitude due to the special properties of the underlying
$psu(2,2|4)$ Lie superalgebra. We would like to stress this fact as commonly Yangian symmetry is
violated by finite length systems, which is \emph{not} the case for our specific representation.
This is of central importance for us, as typical scattering amplitudes are of short `length' with the order
of, say, $n=4,5,6$ particles involved.

\subsection*{Outline}
In section \ref{sect-amplitudes}, after summarising our conventions for writing amplitudes in
on-shell superspace, we review how the conventional and dual superconformal symmetries are realised on
the amplitudes. Then, in section \ref{sect-yangian}, we study the closure of the conventional and
dual superconformal algebras, and find that one obtains a Yangian algebra. Section \ref{sect-conclusions} contains the conclusions and an outlook. The definitions of all generators of the conventional and dual superconformal algebras as well as their commutation relations are collected in an appendix.

\section{Properties of scattering amplitudes}\label{sect-amplitudes}

In this paper we consider colour-ordered partial amplitudes, and we use the spinor-helicity formalism (for a review, see e.g. \cite{Mangano:1990by}),
where the on-shell momenta $p^{\mu}_{i}$ of the $i$'th particle (with $p_{i}^2 = 0$) are written as a product of
commuting spinors,
\begin{equation}
p_i^\mu (\sigma_\mu)^{\alpha \dot{\alpha}} = p_{i}^{\alpha \dot{\alpha}} =  \lambda_i^{\alpha} \tilde{\lambda}_i^{\alpha}\,.
\end{equation}
For contractions between spinors we will use the standard notation $\langle i j \rangle = \lambda^{\alpha}_{i} \lambda_{j \alpha}$.
The spinors $\lambda$ and $\tilde\lambda$ are defined only modulo a complex rescaling $\lambda \rightarrow \alpha \lambda$, $\tilde\lambda \rightarrow \alpha^{-1} \tilde\lambda$. The scaling weight is the helicity and is normalised so that $\lambda$ has helicity $-\tfrac{1}{2}$ and $\tilde\lambda$ has helicity $+\tfrac{1}{2}$.

\subsection{Amplitudes in on-shell superspace and superconformal symmetry}
\label{section-superspace}

Recall that using Grassmann variables $\eta^{A}$, where $A$ is an $SU(4)$ index, we can define a super-wavefunction
\ba \label{super-wave}
  \Phi(p,\eta) &=& G^{+}(p) + \eta^A \Gamma_A(p) + \frac{1}{2}\eta^A \eta^B S_{AB}(p)
  + \frac{1}{3!}\eta^A\eta^B\eta^C \epsilon_{ABCD} \bar\Gamma^{D}(p) \nonumber \\
  &&\  + \frac{1}{4!}\eta^A\eta^B\eta^C \eta^D \epsilon_{ABCD} G^{-}(p)\,,
\ea
which incorporates as its components all on-shell states of $\mathcal{N}=4$ SYM.
For example, $G^{+}(p)$ and $G^{-}(p)$ correspond to helicity plus and helicity minus gluons, respectively,
with ingoing momentum $p^{\mu}$. The Grassmann variables $\eta$ carry helicity $+\tfrac{1}{2}$ so that  the whole multiplet carries helicity $+1$.
Since the $\mathcal{N}=4$ supermultiplet is PCT self-conjugate, we could have equally chosen
an anti-chiral representation (see \cite{dhks5,dhks6} for more explanations).

We can write the amplitudes in the on-shell superspace with coordinates $(\lambda_i,\tilde\lambda_i,\eta_i)$ \cite{Nair:1988bq,Witten:2003nn,Georgiou:2004by},
\be\label{super-amplitude}
{\cal A}_n\big(\lambda_i,\tilde{\lambda}_i,\eta_i \big) = \mathcal{A}\left( \Phi_1 \ldots \Phi_n \right)\,.
\ee
Since the helicity of each supermultiplet $\Phi_i$ is $1$ then the amplitude obeys
\be
\label{helicity}
h_i \mathcal{A}_n(\lambda_i,\tilde\lambda_i,\eta_i) = \mathcal{A}_n(\lambda_i,\tilde\lambda_i,\eta_i)\,,
\ee
where
\be
h_i = -\tfrac{1}{2} \lambda_i^\alpha \frac{\partial}{\partial \lambda_i^{\alpha}} + \tfrac{1}{2} \tilde\lambda_i^{\dot\alpha} \frac{\partial}{\partial \tilde\lambda_i^{\dot\alpha}} + \tfrac{1}{2} \eta_i^A \frac{\partial}{\partial \eta_i^A}
\ee
is the $i$th helicity operator.
The amplitude is a distribution,
\be \label{susy-amplitudes}
\mathcal{A}_n = \frac{\delta^{(4)}(p) \delta^{(8)}(q)}{\langle 12 \rangle \ldots \langle n1 \rangle} \mathcal{P}_n(\lambda_i,\tilde{\lambda}_i,\eta_i )\,,
\ee
where the delta functions are consequences of translation invariance and supersymmetry,
\begin{equation}
p^{\alpha \dot\alpha}=\sum_{i=1}^{n} p_{i}^{\alpha \dot\alpha}\,,\qquad q^{\alpha A}=\sum_{i=1}^{n} q_{i}^{\alpha A}\,,\quad {\rm with}  \quad q_{i}^{\alpha A} = \lambda_{i}^{\alpha} \eta_{i}^{A} \,.
\end{equation}
The function $\mathcal{P}_n( \lambda_i,\tilde{\lambda}_i,\eta_i )$ is a polynomial in the $\eta_{i}$,
with terms of Grassmann degree $4 m$ corresponding to N${}^{m}$MHV amplitudes.
The first term in $\mathcal{P}_n( \lambda_i,\tilde{\lambda}_i,\eta_i )$ is equal to $1$, so that
we recover the well-known formula for MHV amplitudes \cite{Nair:1988bq},
\be
\mathcal{A}_n^{\rm MHV} = \frac{\delta^{(4)}(p) \delta^{(8)}(q)}{\langle 12 \rangle \ldots \langle n1 \rangle} \,.
\ee
An explicit expression for $\mathcal{P}_n( \lambda_i,\tilde{\lambda}_i,\eta_i )$ for all non-MHV tree-level amplitudes can be found in \cite{Drummond:2008cr}.
The function $\mathcal{P}_n( \lambda_i,\tilde{\lambda}_i,\eta_i )$ is constrained by the fact that (at least at tree-level) the amplitude should be
annihilated by all generators of the conventional superconformal algebra
(the explicit form of the generators in the on-shell superspace was given in \cite{Witten:2003nn}).
Let $J^{(0)}$ denote the conventional superconformal
generators satisfying
\begin{equation}\label{level0}
\lbrack J^{(0)}_a , J^{(0)}_b \} = f_{ab}{}^c J^{(0)}_c \,,
\end{equation}
where $f_{ab}{}^{c}$ are the structure constants of the $psu(2,2|4)$ algebra and the mixed brackets $[\cdot , \cdot \}$ denote the graded commutator defined in the usual way, i.e.
\begin{equation}
[ O_{1} , O_{2} \} = O_{1} O_{2} - (-1)^{{\rm deg}(O_{1}) {\rm deg}(O_{2})} O_{2} O_{1} \,.
\end{equation}
Here ${\rm deg}(O)$ is the Grassmann degree of $O$.

The generators $J^{(0)}_a$ are in fact sums of generators acting along the chain of colour-ordered scattered particles,
\begin{equation}
J^{(0)}_a = \sum_{i=1}^{n} J^{(0)}_{i a}\,,
\end{equation}
and we call $J^{(0)}_{i a}$ the density of the generator $J^{(0)}_{a}$.
In particular we have the second order constraint,
\be \label{kannihilates}
k_{\alpha \dot \alpha} \, \mathcal{A}_n = 0\,, \qquad k_{\alpha \dot \alpha} =  \sum_{i=1}^{n} k_{i \alpha \dot \alpha} = \sum_{i=1}^{n}  \frac{\partial}{\partial \lambda_i^{\alpha}}\frac{\partial}{\partial \tilde\lambda_i^{\dot\alpha}}\,,
\ee
where $k_{\alpha \dot \alpha}$ is the generator of special conformal transformations.
The expressions for the remaining generators of the conventional superconformal algebra are given in the appendix.
We will use the superscript zero for the level zero generators to distinguish them from higher level generators to be defined below.
In some formulae, like in (\ref{kannihilates}), this superscript will be omitted for the level zero generators for the sake of legibility.

We can trivially extend $psu(2,2|4)$ to $su(2,2|4)$ by including the central charge $c$ which is
related to the total helicity via $c = \sum_{i=1}^{n} (1-h_i)$ and therefore annihilates the amplitudes, as can be seen
from equation (\ref{helicity}).
In fact this is also true for the density $c_{i}$ which is related to the helicity generators via $c_i = 1-h_i$.
We can further extend to $u(2,2|4)$ by including the hypercharge $b$, but we must remember that this is not a symmetry of the amplitudes.
Thus we can take the structure constants $f_{ab}{}^c$ to be those of $u(2,2|4)$.

\subsection{Dual superconformal symmetry}

It has become clear recently that amplitudes have an additional symmetry, dual superconformal symmetry
 \cite{dhks5}.
This symmetry is most conveniently seen by introducing dual variables $x_i$ and $\theta_{i}$,
\be \label{defdualvars}
(x_i-x_{i+1})_{\a \dot\alpha}  - \lambda_{i\, \a}\, \tilde{\lambda}_{i\, \dot\alpha} = 0, \qquad (\theta_i - \theta_{i+1})_\alpha^A - \lambda_{i \alpha} \eta_i^A = 0\,.
\ee
For now, we will identify the dual superspace points $(x_{n+1},\theta_{n+1})$ and $(x_{1},\theta_{1})$,
as suggested by momentum conservation and supersymmetry.
However, we will come back to this issue momentarily and explain why it is more appropriate to consider $(x_{n+1},\theta_{n+1}) \neq (x_{1},\theta_{1})$.

First we recall how the generators of the dual superconformal algebra are constructed \cite{dhks5}.
We begin with the dual coordinates $x_i$ and $\theta_i$. On this chiral superspace, the conformal generator reads
\be
K^{\alpha \dot{\alpha}} = \sum_{i=1}^{n} \biggl[ x_i^{\alpha \dot{\beta}} x_i^{ \dot{\alpha} \beta} \frac{\partial}{\partial x_i^{\beta \dot{\beta}}} + x_i^{\dot{\alpha}\beta} \theta_i^{\alpha B} \frac{\partial}{\partial \theta_i^{\beta B}} \biggr]\,.
\label{Kdualcoords}
\ee
The action of the dual conformal generator on the variables $(\lambda_i^\alpha,\tilde\lambda_i^{\dot\alpha},\eta_i^A)$ is
found by extending (\ref{Kdualcoords}) so that it commutes with the constraints (\ref{defdualvars}) modulo constraints. In other words it is constructed so that it preserves the surface defined by (\ref{defdualvars}). There is a choice in how to do this related to how the conformal transformation of $\lambda_i$ is split up between the points $x_i$ and $x_{i+1}$. The choice we will make (as in \cite{dhks5}) associates $\lambda_i$ entirely with the point $x_i$ and $\tilde\lambda_i$ with the point $x_{i+1}$,
\be
K^{\alpha \dot{\alpha}} = \sum_{i=1}^{n} \biggl[ x_i^{\alpha \dot{\beta}} x_i^{\dot{\alpha} \beta} \frac{\partial}{\partial x_i^{\beta \dot{\beta}}}
+ x_i^{\dot{\alpha} \beta}  \theta_i^{\alpha B} \frac{\partial}{\partial \theta_i^{\beta B}} +
  x_i^{\dot{\alpha} \beta} \lambda_i^{\alpha} \frac{\partial}{\partial \lambda_i^{\beta}}
  + x_{i+1}^{\alpha \dot{\beta}} \tilde{\lambda}_i^{\dot{\alpha}}
  \frac{\partial}{\partial \tilde{\lambda}_{i}^{\dot{\beta}}} + \tilde{\lambda}_i^{\dot{\alpha}} \theta_{i+1}^{\alpha B} \frac{\partial}{\partial \eta_i^B} \biggr]\,.
\label{Kfullform}
\ee
We can obtain all generators of the dual superconformal algebra in the same way. For our purposes it will be useful to recall here the form of the superconformal generator $S_\alpha^A$,
\be
S_{\alpha}^A = \sum_{i=1}^{n} \biggl[ -\theta_{i \alpha}^{B} \theta_i^{\beta A}
  \frac{\partial}{\partial \theta_i^{\beta B}} + x_{i \alpha}{}^{\dot{\beta}} \theta_i^{\beta
    A} \frac{\partial}{\partial x_i^{\beta \dot{\beta}}} + \lambda_{i \alpha}
  \theta_{i}^{\gamma A} \frac{\partial}{\partial \lambda_i^{\gamma}} + x_{i+1\,
    \alpha}{}^{\dot{\beta}} \eta_i^A \frac{\partial}{\partial \tilde{\lambda}_i^{\dot{\beta}}} -
  \theta_{i+1\, \alpha}^B \eta_i^A \frac{\partial}{\partial \eta_i^{B}} \biggr] \,.
\label{Sfullform}
\ee
The expressions for the remaining generators of the dual superconformal algebra are given in the appendix.
It will be sufficient for us to consider $K^{\alpha \dot{\alpha}}$ and $S_{\alpha}^A$ since the other
generators of the dual superconformal algebra can be seen to be trivially related to generators of
the conventional superconformal algebra.

Recall \cite{dhks5} that we can act on the amplitude as a distribution by introducing an additional point $(x_{n+1},\theta_{n+1}) \neq(x_1,\theta_1)$.
Then the terms in the generators which depend on $\partial/\partial x_i$ or $\partial / \partial \theta_i$ (i.e. the first two terms in (\ref{Kfullform}) and (\ref{Sfullform})) are summed up to $n+1$.
{}From the dual space perspective this means that the chain of points no longer forms a closed loop but rather an open curve.
The delta functions  $\delta^{(4)}(p) \delta^{(8)}(q)$
appearing in a generic super-amplitude, c.f. (\ref{susy-amplitudes}), can then be written as $\delta^{(4)}(x_1-x_{n+1}) \delta^{(8)}(\theta_1 - \theta_{n+1})$
and hence reimposes the closure of the loop which is thus a consequence of translation invariance and supersymmetry.

The product $\delta^{(4)}(p) \delta^{(8)}(q)=\delta^{(4)}(x_1-x_{n+1})\delta^{(8)}(\theta_{1}-\theta_{n+1})$ can be seen to be dual superconformally invariant \cite{dhks5}.
The full amplitude is dual superconformally covariant, with the conformal weights coming exclusively from the denominator $\langle 12 \rangle \ldots \langle n 1 \rangle$,
\be \label{dualKcovariance}
K^{\alpha \dot \alpha} \mathcal{A}_n = - \sum_{i=1}^n x_i^{\alpha \dot \alpha} \mathcal{A}_n\,, \hspace{30pt} S_\alpha^A \mathcal{A}_n = - \sum_{i=1}^n \theta_{i \alpha}^A \mathcal{A}_n \,.
\ee
The operators which annihilate the amplitude are therefore
\be
\tilde{K}^{\alpha \dot\alpha}=K^{\alpha \dot\alpha} + \sum_{i=1}^n x_i^{\alpha \dot\alpha} \hspace{10pt} \text{ and } \hspace{10pt} \tilde{S}_{\alpha}^{A} = S_{\alpha}^A + \sum_{i=1}^n \theta_{i \alpha}^A\,.
\label{tildeKS}
\ee
We want to consider the commutation of the charges which annihilate the amplitude in both the conventional and dual superconformal algebras. To do this we must bring $\tilde K$ and $\tilde S$ to a form where they only act on the on-shell superspace coordinates  $(\lambda_{i},\tilde{\lambda}_{i},\eta_{i})$. First we will use the constraints (\ref{defdualvars}) to eliminate the dual variables $x_i$ and $\theta_i$ for $2\leq i \leq n+1$ in favour of the on-shell superspace coordinates,
\be
x_i^{\alpha \dot\alpha} = x_1^{\alpha \dot\alpha} - \sum_{j<i} \lambda_j^{\alpha}\tilde \lambda_j^{\dot\alpha}\,, \qquad \theta_i^{\alpha A} = \theta_1^{\alpha A} - \sum_{j<i} \lambda_j^{\alpha} \eta_j^A\, \hspace{10pt} \text{ for } \hspace{10pt} 2\leq i \leq n+1
\,,
\ee
so that the terms acting on the eliminated variables can be dropped. The terms which act on $x_1$ and $\theta_1$ can also be dropped because, once $x_2,\ldots, x_{n+1}$ and $\theta_2,\ldots, \theta_{n+1}$ have been eliminated, the dual translation invariance and dual supersymmetry of the amplitude simply state the independence of the amplitude on $x_1$ and $\theta_1$,
\be
P_{\alpha \dot\alpha} \mathcal{A}_n = \frac{\partial}{\partial x_1^{\alpha \dot\alpha}} \mathcal{A}_n = 0, \hspace{30pt} Q_{\alpha A}  \mathcal{A}_n = \frac{\partial}{\partial \theta^{\alpha A}} \mathcal{A}_n = 0\,.
\ee
Indeed we know from the start that the amplitude can be written so that it only depends on the on-shell superspace coordinates $(\lambda_{i},\tilde{\lambda}_{i},\eta_{i})$. Thus we keep only the final three terms in both (\ref{Kfullform}) and (\ref{Sfullform}).

Then if we inspect the remaining terms in $\tilde K$ and $\tilde S$ in equation (\ref{tildeKS}) which depend on $x_1$ and $\theta_1$ we see that they annihilate the amplitude by themselves. For example, if we look at $\tilde S$ we find the following terms which depend on $x_1$ and $\theta_1$,
\begin{align}
&\sum_{i=1}^n \Bigl[ \theta_1^{\gamma A} \lambda_{i \alpha} \frac{\partial}{\partial \lambda_i^\gamma} - \theta_{1\alpha}^B \eta_i^A \frac{\partial}{\partial \eta_i^B} +x_{1 \alpha}{}^{\dot\beta} \eta_i^A \frac{\partial}{\partial \tilde{\lambda}_i^{\dot \beta}} - \theta_{1 \alpha}^A \Bigr] =\notag\\
&{}\hspace{4cm}= \theta_{1 \gamma}^B(-\delta_B^A m^{\gamma}{}_{\alpha} + \tfrac{1}{2}\delta_\alpha^\gamma \delta_B^A (d+c) + \delta_\alpha^\gamma r^{A}{}_{B}) +x_{1 \alpha}{}^{\dot\beta} \bar{q}_{\dot \beta}^A\,.
\label{extraterms}
\end{align}
Here we use the notation $m,d,c,r,\bar{q}$ for the generators of the superconformal algebra whose explicit forms are given in the appendix.
Since all terms on the r.h.s. of (\ref{extraterms}) are proportional to generators of the original superconformal algebra, they annihilate the amplitude. The remaining terms in $\tilde K$ and $\tilde S$ contain only the on-shell superspace coordinates $(\lambda_{i},\tilde{\lambda}_{i},\eta_{i})$ and derivatives with respect to them. We will call these symmetry generators $K'$ and $S'$,
\begin{align}
K' &= -  \sum_{i=1}^{n} \left\lbrack \sum_{j=1}^{i-1} \lambda_{j}^{\beta} \tilde{\lambda}_j^{\dot{\alpha}}  \lambda_i^{\alpha} \frac{\partial}{\partial \lambda_i^{\beta}}
  + \sum_{j=1}^{i} \lambda_{j}^{\alpha} \tilde{\lambda}_{j}^{\dot{\beta}} \tilde{\lambda}_i^{\dot{\alpha}}
  \frac{\partial}{\partial \tilde{\lambda}_i^{\dot{\beta}}} + \sum_{j=1}^{i} \tilde{\lambda}_i^{\dot{\alpha}} \lambda_{j}^{\alpha} \eta_{j}^{B} \frac{\partial}{\partial \eta_i^{B}} + \sum_{j=1}^{i-1} \lambda_j^{\alpha} \tilde{\lambda}_{j}^{\dot\alpha} \right\rbrack \,, \label{Kprime}\\
S^{'A}_{\alpha} &= -\sum_{i=1}^{n} \left\lbrack \sum_{j=1}^{i-1}  \lambda_{j}^{\gamma} \eta_{j}^{A} \lambda_{i \alpha} \frac{\partial}{\partial \lambda_i^{\gamma}} + \sum_{j=1}^{i} \lambda_{j \alpha} \tilde{\lambda}_{j}^{\dot{\beta}} \eta_{i}^{A} \frac{\partial}{\partial \tilde{\lambda}_i^{\dot{\beta}}} - \sum_{j=1}^{i} \lambda_{j \alpha} \eta_{j}^{B} \eta_{i}^{A} \frac{\partial}{\partial \eta_i^{B}} + \sum_{j=1}^{i-1} \lambda_{j \alpha} \eta_{j}^{A} \right\rbrack \,.  \label{Sprime}
\end{align}
That these operators are symmetries is the expression of dual superconformal symmetry purely in terms of the data defining the scattering amplitudes.

We could equivalently have arrived at the form of $K'$ and $S'$ by noting that $\tilde K$ and $\tilde S$ commute with
the dual translation generator $P_{\alpha \dot\alpha}$ and the dual supersymmetry generator $Q_{\alpha A}$
 modulo terms which themselves annihilate the amplitude. Therefore we could simply use the latter to set $x_1$ and $\theta_1$ to zero in (\ref{tildeKS}),
immediately leading to (\ref{Kprime}) and (\ref{Sprime}).

\section{Yangian symmetry}\label{sect-yangian}

What happens when we commute the generators of the conventional
and dual superconformal symmetries to obtain higher charges which are realised on the amplitudes?
The answer is that the closure of the two symmetry algebras defines a Yangian algebra.
We will show presently that the dual superconformal generators are equivalent to level one
generators $J^{(1)}$ that are symmetries of the amplitudes and
that satisfy
\begin{equation}\label{level01}
\lbrack J^{(1)}_a , J^{(0)}_b \} = f_{ab}{}^c J^{(1)}_c\,.
\end{equation}
The level one generators are defined by
\be
\label{yangianspinchain}
J^{(1)}_{a} =  f_{a}{}^{cb} \sum_{1\leq i < j \leq n}  J^{(0)}_{i b} J^{(0)}_{j c}\,,
\ee
where $f^{cb}{}_{a}$ is obtained from $f_{ab}{}^c$ by raising and lowering indices
with the metric $g_{ab}$ of the algebra. 
Further they satisfy the following relation (Serre relation)
\begin{align}
&[J^{(1)}_a , [J^{(1)}_b,J^{(0)}_c\} \} + (-1)^{|a|(|b| + |c|)} [J^{(1)}_b,[J^{(1)}_c,J^{(0)}_a\} \} + (-1)^{|c|(|a|+|b|)} [J^{(1)}_c,[J^{(1)}_a,J^{(0)}_b\} \} \notag \\
&= h (-1)^{|r||m|+|t||n|}\{J^{(0)}_l,J^{(0)}_m,J^{(0)}_n] f_{ar}{}^{l} f_{bs}{}^{m} f_{ct}{}^{n} f^{rst}.
\end{align}
Here we use the mixed brackets $[\cdot , \cdot \}$ to denote the graded commutator, as before,  and the symbol $\{ \cdot , \cdot , \cdot ]$ to denote the graded symmetriser. The index gradings denote the Grassmann degree of the corresponding generators, i.e. $|a| = {\rm deg}(J_a)$ etc. The constant $h$ is dependent on conventions.

The conditions under which (\ref{yangianspinchain}) is sufficient to give a representation of the Yangian
(i.e. that the level one generators obey the Serre relation) were discussed in detail in \cite{Dolan:2004ps}.
A sufficient condition is that the adjoint appears only once in the tensor product of the representation of the single-site level-zero generators $J_{ia}^{(0)}$ with its conjugate. The representation of interest here, namely the on-shell gluon supermultiplet, does satisfy this property \cite{Dolan:2004ps}.
Therefore to show that the tree-level amplitudes have a Yangian symmetry it will be sufficient to show that dual superconformal symmetry implies invariance under the generators $J^{(1)}_a$ given in (\ref{yangianspinchain}).

Let us start with constructing the level one supersymmetry generator $q^{(1)}{}_\alpha^A$. Looking at formula (\ref{yangianspinchain}) we expect to find terms of the form $m_i q_j$, $d_i q_j$, $p_i \bar{s}_j$, $q_i r_j$ and $c_i q_j$ (all antisymmetrised on $i$ and $j$). Recall that here we are using the notation $m_i$ for the density of the Lorentz generator $m$ and similarly for the other generators of the superconformal algebra whose explicit forms are given in the appendix. The central charge density $c_i$ appears because certain elements of the metric are off-diagonal. In particular we have $g_{sq}\neq 0$, $g_{bc}\neq 0$, $f_{sb}{}^{s}\neq 0$ and hence $0\neq f^{qc}{}_{q} = - f_q{}^{qc}$.
Since $c$ is central $b$ will never appear on the r.h.s. of (\ref{yangianspinchain}).

We would like to show that the dual superconformal symmetry generator $S_{\alpha}^{A}$ is related to the level one supersymmetry generator $q^{{(1)}A}_\alpha$.
To see that this is the right way to attempt to identify dual superconformal symmetry with the Yangian symmetry, it suffices to note that this is the only
generator of $u(2,2|4)$ having the same labels as $S_{\alpha}^{A}$, namely a chiral Lorentz spinor index and an upstairs $su(4)$ index.

We know that dual superconformal symmetry implies that the amplitude is annihilated by the operator $S'$, given in equation (\ref{Sprime}).
Looking at (\ref{yangianspinchain}) we note that $f{}_{a}{}^{cb}$ has the symmetry property $f{}_{a}{}^{cb}=- (-1)^{|b| |c|}  f{}_{a}{}^{bc}$, 
where $|a|$ is zero (one) for $a$ corresponding to a bosonic (fermionic) generator. 
We can match this symmetry property by anti-symmetrising
the indices $i$ and $j$ under the sum in (\ref{Sprime}).
This can be achieved by adding to $S_{\alpha}^{'A}$ the term
\begin{equation}\label{modificationS}
\Delta S_{\alpha}^{A} = \frac{1}{2} \left\lbrack - q^{A}_{\gamma} m^{\gamma}_{\alpha} + q^{A}_{\alpha} \frac{1}{2} d_{\lambda} + n q^{A}_{\alpha} + p_{\alpha}^{\dot{\beta}} \bar{s}^{A}_{\dot{\beta}} + q^{B}_{\alpha} r^{A}_{B} - q^{A}_{\alpha} \frac{1}{4} d_{\eta}
\right\rbrack \,,
\end{equation}
where $d_{\lambda} = \sum_i \lambda_{i}^{\beta} \frac{\partial}{\partial \lambda_{i}^{\beta}}$ and $d_{\eta}=\sum_i \eta_{i}^{B} \frac{\partial}{\partial \eta_{i}^{B}}$
are counting operators. In (\ref{modificationS}) they always appear multiplied
by $q^{A}_{\alpha}$ such that all operators in (\ref{modificationS}) annihilate the amplitudes, and hence we
have the freedom of modifying $S_{\alpha}^{'A}$ in this way.

After a short calculation
we obtain
\begin{eqnarray}\label{modifiedS}
S_{\alpha}^{'A} + \Delta S_{\alpha}^{A} &=& \frac{1}{2} \sum_{i>j}  \left\lbrack
m_{i \alpha}^{\gamma} q_{j \gamma}^{A}-\frac{1}{2} (d_{i}+c_{i}) q_{j \alpha}^{A} + p_{i \alpha}^{\dot{\beta}} \bar{s}_{j \dot{\beta}}^{A} + q_{i \alpha}^{B} r^{A}_{j B}   - ( i \leftrightarrow j ) \right\rbrack  \nonumber \\
&& + \sum_{i=1}^{n} q_{i \alpha }^{A} c_{i} - \frac{1}{2} q_{\alpha}^{A}\,.
\end{eqnarray}
We will now argue that the terms in the second line of (\ref{modifiedS}) annihilate the
scattering amplitudes on their own and we may therefore drop them.
Indeed, for the second term this is obvious, and from the discussion in section \ref{sect-amplitudes}
we already know that not only $c$, but also the density $c_{i}$ annihilates the amplitudes.
So we can make the definition
\begin{equation}\label{q1level1final}
q^{(1)A}_{\alpha} := \sum_{i>j}  \left\lbrack
m_{i \alpha}^{\gamma} q_{j \gamma}^{A}-\frac{1}{2} (d_{i}+c_{i}) q_{j \alpha}^{A} + p_{i \alpha}^{\dot{\beta}} \bar{s}_{j \dot{\beta}}^{A} + q_{i \alpha}^{B} r^{A}_{j B}  - ( i \leftrightarrow j )  \right\rbrack \,.
\end{equation}
Here we have included an extra overall factor of 2 for convenience.
By construction $q^{(1)A}_{\alpha}$ annihilates all tree level scattering amplitudes,
and it contains as an essential part the dual generator $S_{\alpha}^{'A}$.

We will now show that (\ref{q1level1final}) is exactly of the form (\ref{yangianspinchain}).
This can be done by computing the structure constants $f_{a}{}^{cb}$ in (\ref{yangianspinchain})
for $a$ corresponding to $q^{A}_{\alpha}$. We do this explicitly in the appendix (see equations (\ref{J1formula},\ref{q1calc1},\ref{q1calc2})).
Here we prefer to give a simpler derivation by noting that we have found precisely the types of terms we expected from (\ref{yangianspinchain}). Therefore we only
need to check that the relative coefficients in (\ref{q1level1final}) are
correct. We will do this by verifying the algebra relations (\ref{level01}).
To begin with, the equations
\begin{equation}
\{ q^{(1)A}_{\alpha} , \bar{s}^{B}_{\dot{\alpha}} \} = 0\,,\qquad \lbrack q^{(1) A}_{\alpha} , p^{ \beta \dot{\beta} } \rbrack = 0\,,
\end{equation}
fix all the relative coefficients in (\ref{q1level1final}) to take the values given there, except for the $c q$ term.
Since $c$ is a central charge it does not appear on the l.h.s. of
any commutation relation. Nevertheless, we can check its coefficient in (\ref{q1level1final})
by verifying the algebra relations (\ref{level01}), as we will see presently.
In order to do this, we compute
\begin{equation}
\{  q^{(1)A}_{\alpha} , \bar{q}^{}_{\dot{\alpha} B} \} =: \delta^{A}_{B} p_{\alpha \dot{\alpha}}^{(1)}\,,
\end{equation}
and obtain
\begin{equation}
p_{\alpha \dot{\alpha}}^{(1)} =  \sum_{i>j}  \left\lbrack \left( m_{i \alpha}^{\gamma} \delta_{\dot{\alpha}}^{\dot{\gamma}} + \bar{m}_{i \dot{\alpha}}^{\dot{\gamma}} \delta_{\alpha}^{\gamma} - d_{i} \delta_{\alpha}^{\gamma} \delta_{\dot{\alpha}}^{\dot{\gamma}} \right) p_{j \gamma \dot{\gamma}} + \bar{q}_{i \dot{\alpha} C } q_{j \alpha}^{C}  - ( i \leftrightarrow j ) \right\rbrack \,.
\end{equation}
Now we are in a position to verify selfconsistently that
\begin{equation}\label{check3}
\lbrack p_{\alpha \dot{\alpha}}^{(1)} , \bar{s}^{\dot{\beta}}_{A} \rbrack = \delta_{\dot\alpha}^{\dot\beta} {q}_{\alpha}^{(1) A}\,.
\end{equation}
Indeed, the ${q}_{\alpha}^{(1) A}$ we find from (\ref{check3}) is exactly (\ref{q1level1final}), which confirms the coefficient of the $cq$ term there.

This completes the proof that (\ref{q1level1final})  is equivalent to the corresponding case in (\ref{yangianspinchain}), up
to the overall normalisation.
We remark that the structure of the $su(2,2|4)$ algebra is such that, given $q^{(1)}$, we can obtain all other level one generators from the relations (\ref{level01}). From \cite{Dolan:2003uh,Dolan:2004ps} we then
know that (\ref{level01})
and the Serre relations are satisfied, such that the symmetry operators of the amplitudes form
a Yangian algebra.
We remark that $p^{(1)}$ is related to the dual generator $K^{'}$ in a way
very similar to how $q^{(1)A}$ is related to $S^{'A}$. Therefore no
further new generators appear from commuting the Yangian generators
with $K^{'}$.

Before concluding we give a convenient (super-)matrix form of
the Yangian generators.
Introducing super-indices $\uu{A} = (\alpha, \dot{\alpha}, A)$ we can define
 $(4|4)\times (4|4)$ super-matrices $J_{i}{}^{\uu{A}}{}_{\uu{B}}$ satisfying the $u(2,2|4)$
 (anti)commutation relations (for $i=j$)
\begin{equation}
\label{supercom}
[  J_{i}{}^{\uu A}{}_{\uu B}\, , \,  J_{j}{}^{\uu C}{}_{\uu D} \} = \delta_{ij} \left\lbrack
\delta^{\uu C}_{\uu B}\,  J_{i}{}^{\uu A}{}_{\uu D}- (-1)^{(|\uu A|+|\uu B|)(\uu C|+|\uu D|)} \delta^{\uu A}_{\uu D}\,  J_{i}{}^{\uu C}{}_{\uu B} \right\rbrack\,,
\end{equation}
where we have introduced the index-gradings $|\alpha|=|\dot\alpha|=0$ and $|A|=1$. 
One can show that the definition
\begin{eqnarray}\label{JABdef2}
J_{i}{}{}^{\uu{A}}{}_{\uu{B}} &=& \left (
\begin{matrix}
\partial_{i}{}^{\alpha} \lambda_{i}{}_{\beta} & \partial_{i}{}^{\alpha} \partial_{i}{}_{\dot{\beta}} & \partial_{i}{}^{\alpha} \partial_{i}{}_{B}  \\
\tilde{\lambda}_{i}{}^{\dot{\alpha}} \lambda_{i}{}_{\beta} & \tilde{\lambda}_{i}{}^{\dot{\alpha}} \partial_{i}{}_{\dot{\beta}} & \tilde{\lambda}_{i}{}^{\dot{\alpha}} \partial_{i}{}_{B} \\
\eta_{i}{}^{A} \lambda_{i}{}_{\beta} & \eta_{i}{}^{A} \partial_{i}{}_{\dot{\beta}} & \eta_{i}{}^{A} \partial_{i}{}_{B}  \end{matrix}
\right ) \\
&=&
\left (
\begin{matrix}
m^\alpha{}_\beta - \frac{1}{2}\, \delta^\alpha_\beta\, (d+\frac{1}{2}c-\frac{1}{4}t) & k^{\alpha}{}_{\dot\beta} & s^{\alpha}{}_{B}  \\
p^{\dot{\alpha}}{}_{\beta} & \overline{m}^{\dot \alpha}{}_{\dot\beta} + \frac{1}{2}\, \delta^{\dot\alpha}_{\dot\beta}\, (d-\frac{1}{2}c+\frac{1}{4}t)
& {\bar q}^{\dot\alpha}{}_B \\
q^A{}_\beta & {\bar s}^A{}_{\dot\beta} & -r^A{}_B-\frac{1}{2}\delta^A_B\, (\frac{1}{2}c+\frac{1}{4}t)
\end{matrix}
\right )_{i} \nonumber
\end{eqnarray}
satisfies (\ref{supercom}). Here we used a shorthand notation for the derivatives, c.f. (\ref{shortderiv}), and
$t_{i}$ is the super-trace of the matrix. It can be identified with the hypercharge $b_{i}$, but the definition of $b_{i}$ is arbitrary in that one can add an arbitrary amount of the central charge and a constant. So the general expression for the hypercharge is $b_{i}  = t_{i} + n_1 c_{i} + n_2$.
The explicit expression for $t_{i}$ in (\ref{JABdef2}) is $t_{i}=2-\lambda_{i}{}^{\gamma}\partial_{i}{}_{\gamma}+\tilde{\lambda}_{i}{}^{\dot{\gamma}}\partial_{i}{}_{\dot{\gamma}}-\eta_{i}{}^{C}\partial_{i}{}_{C}$.

We can now write down the level one generators using the super-matrix notation.
They take the form
\begin{equation}\label{level1-matrix}
 J^{(1)\, \uu A}{}_{\uu B} := - \sum_{i>j} (-1)^{|\uu{C}|} ( J_{i}^{\uu A}{}_{\uu C}\, J_j^{\uu C}{}_{\uu B}
-\, J_j^{ \uu A}{}_{\uu C} \, J_i^{ \uu C}{}_{\uu B}  )\,,
\end{equation}
which can be seen to be equivalent to (\ref{yangianspinchain}).
An advantage of the (super-)matrix formulation is that using (\ref{supercom}) it is straightforward to compute (anti-)commutators involving Yangian
generators. For example, we can easily verify that (\ref{level1-matrix}) obeys the Yangian relation (\ref{level01}),
\begin{equation}
[  J^{(1)\, \uu A}{}_{\uu B}\, , \,  J^{\uu C}{}_{\uu D} \}
= \delta^{\uu C}_{\uu B}\,  J^{(1)\, \uu A}{}_{\uu D}- (-1)^{(|\uu A|+|\uu B|)(\uu C|+|\uu D|)} \delta^{\uu A}_{\uu D}\, J^{(1)\,\uu C}{}_{\uu B}\,.
\end{equation}
Let us comment on the consistency of the level one Yangian generators defined by (\ref{yangianspinchain})
with the cyclicity of the scattering amplitudes\footnote{We thank Edward Witten for drawing our attention to this point.}.
Let us consider the definition we would obtain if we rotated the chain of points cyclically by one step,
\be
{\tilde J}^{(1)}_a = f_{a}{}^{cb} \sum_{2\leq i<j \leq n+1} J_{ib} J_{jc}\,.
\ee
Here we identify $J_{n+1 \, a}$ with $J_{1a}$. Since the amplitudes are cyclic, both ${J}^{(1)}_a$ and ${\tilde J}^{(1)}_a$  should annihilate the amplitudes and hence so should their difference.
In the super-matrix notation it is easy to evaluate this difference,
\begin{eqnarray}\label{cycl-analysis}
J^{(1)\, \uu A}{}_{\uu B} - \tilde{J}^{(1)\, \uu A}{}_{\uu B} &=& - 2\, \sum_{k=1}^{n} (-1)^{|\uu{C}|}
( J_{k}^{\uu A}{}_{\uu C}\, J_1^{\uu C}{}_{\uu B}
-\, J_1^{ \uu A}{}_{\uu C} \, J_k^{ \uu C}{}_{\uu B}  ) \,.
 \end{eqnarray}
Here $\sum_{k=1}^{n} J_{k}{}^{\uu{A}}{}_{\uu{C}} = J{}^{\uu{A}}{}_{\uu{C}}$ is itself a symmetry of the amplitudes, and therefore the r.h.s. of (\ref{cycl-analysis}) annihilates the amplitudes up to
 a term proportional to $(-1)^{|\uu{C}|} [ J_{1}{}^{\uu{A}}{}_{\uu{C}}, J_{1}{}^{\uu{C}}{}_{\uu{B}} \}$. The latter may be computed
using (\ref{supercom}), yielding
\begin{eqnarray}
(-1)^{|\uu{C}|} [ J_{1}{}^{\uu{A}}{}_{\uu{C}}, J_{1}{}^{\uu{C}}{}_{\uu{B}} \} &=&  (-1)^{|\uu{C}|} ( \delta^{\uu{C}}_{\uu{C}} J_{1}{}^{\uu{A}}{}_{\uu{B}} - (-1)^{(|\uu{A}|+|\uu{C}|)(|\uu{B}|+|\uu{C}|)}  \delta^{\uu{A}}_{\uu{B}} J_{1}{}^{\uu{C}}{}_{\uu{C}} )  \nonumber \\
 &=& \delta^{\uu{A}}_{\uu{B}} \, J_{1}{}^{\uu{C}}{}_{\uu{C}} = \delta^{\uu{A}}_{\uu{B}} \, c_{1} \,,
\end{eqnarray}
where we have used the fact that, importantly, $(-1)^{|\uu{C}|} \delta^{\uu{C}}_{\uu{C}} = 0$ for $u(2,2|4)$ and that the trace of $J_{1}$ is equal to the central charge density $c_{1}$, see (\ref{JABdef2}).
Since the central charge densities $c_{i}$ annihilate the amplitudes we conclude that so does
$J^{(1)\, \uu A}{}_{\uu B} - \tilde{J}^{(1)\, \uu A}{}_{\uu B}$
and therefore the definition (\ref{level1-matrix}) is consistent with
cyclicity of the amplitudes. Note that this emergence of the central charge density is a peculiar
property of $u(m|m)$ superalgebras.

More generally one can say that, up to a term which is proportional to $J_{a}$, the difference between $J^{(1)}_a$ and $\tilde{J}^{(1)}_a$ is given by
\be
f_{a}{}^{cb} f_{bc}{}^{d} J_{1d}\,.
\ee
For general (super)algebras this is proportional to $J_{1a}$ which is not a symmetry generator. However for certain superalgebras the constant of proportionality is zero.
As we show in equations (\ref{ffvanishing})-(\ref{killingform}) in the appendix, this is the case for the algebras with vanishing Killing form which are $psl(n|n),osp(2n+2|2),D(2,1;\alpha),P(n),Q(n)$ \cite{Kac:1977em,Frappat:1996pb}. So the bilocal formula (\ref{yangianspinchain}) is consistent with the cyclicity of the chain for these superalgebras only.

\section{Conclusions and outlook}\label{sect-conclusions}
In this paper we showed that the dual superconformal symmetry found in \cite{dhks5}, together with
the conventional superconformal symmetry of $\mathcal{N}=4$ SYM forms a Yangian symmetry.
The Yangian generators acting on the scattering amplitudes have exactly the same structure as
those relevant for the spectrum of anomalous dimensions of $\mathcal{N}=4$ SYM, at least at tree level.

We conclude with some remarks and give an outlook.

In \cite{Witten:2003nn} it was shown that tree-level amplitudes have remarkable properties when written in (super)twistor space. To obtain the amplitudes in this space, one usually performs a Fourier transform with
respect to the variables $\tilde{\lambda}_i$ and $\eta_i$. Here we will take a slightly different definition
and take a Fourier transform with respect to $\lambda_{i}$, i.e.\be
\tilde{\mathcal{A}}(\mu_i,\tilde{\lambda}_i,\eta_i)  = \int \Bigl(\prod_i d^2 {\lambda}_i  e^{i {\lambda}_i^{{\alpha}} \mu_{i {\alpha}} }\Bigr) \mathcal{A}(\lambda_i,\tilde{\lambda}_i,\eta_i)\,.
\ee
We can think about the generators of the Yangian symmetry written in supertwistor space. If we do this then the level zero generators $J_a^{(0)}$ all become first order differential operators. It is then clear from the formula for the level one generators (\ref{yangianspinchain}) that the latter are given by second order operators. Higher charges generated by commutation of the $J^{(1)}_a$ will result in yet higher order operators.
Let us illustrate the above statements by explicit formulae.
The pairs $(-i \, \mu_{i}^{\alpha} , \tilde{\lambda}_{i}^{\dot{\alpha}})$ form homogeneous coordinates $Z_{i}^{A'}$ on twistor space. Combining the twistors $Z_{i}^{A'}$ with the Grassmann coordinates $\eta_{i}^{A}$ we obtain supertwistors $\mathcal{Z}_{i}^{\uu{A}} = (Z_{i}^{A'},\eta_{i}^{A})$. The level zero generators of $u(2,2|4)$ take a particularly simple form in this language,
\be
J^{(0)}{}^{\uu{A}}{}_{\uu{B}} = \sum_i \mathcal{Z}_{i}^{\uu{A}} \frac{\partial}{\partial \mathcal{Z}_{i}^{\uu{B}}}\,.
\ee
The level one generators are then obviously second order operators acting on the $\mathcal{Z}_i$.
They can be written as
\be
J^{(1)}{}^{\uu{A}}{}_{\uu{B}} = - \sum_{i>j} \left[ \mathcal{Z}_{i}^{\uu{A}} \mathcal{Z}_{j}^{\uu{C}} \frac{\partial}{\partial \mathcal{Z}_{i}^{\uu{C}}} \frac{\partial}{\partial \mathcal{Z}_{j}^{\uu{B}}} - (i \leftrightarrow j) \right] \,.
\ee
It is easy to write down similar formulae for higher level generators.
An interesting absence from this set of generators are order zero operators, i.e. multiplication operators.
These would be the relevant operators for describing the coplanarity/collinearity properties found in \cite{Witten:2003nn}.

Perhaps the most urgent question concerns the fate of the symmetries of the amplitudes
at loop level. Beyond tree level, infrared divergences appear, and the necessary regularisation
{\it a priori} breaks the Yangian symmetry. Nonetheless, in analogy with the case of the spectrum
of anomalous dimensions of $\mathcal{N}=4$ SYM, one might expect that the Yangian symmetry
is realised at loop level as well,
where (at least some of) the Yangian generators receive coupling-dependent deformations.
For example, the dual conformal symmetry is expected to be broken in a way governed by
an anomalous dual conformal Ward identity \cite{Drummond:2007cf,Drummond:2007au}.
Although this Ward identity was initially derived for Wilson loops dual to MHV amplitudes,
it was found in \cite{dhks5,dhks6} that the same Ward identity also holds for one-loop NMHV amplitudes,
and it was conjectured that it should hold for all amplitudes, MHV and non-MHV.
Therefore it seems natural to modify $K'_{\alpha \dot{\alpha}}$ by a coupling-dependent
piece which accounts for this anomaly.
It would be very interesting to find out whether a `deformed' Yangian symmetry is realised at loop level.

Assuming that the amplitudes at loop level have a `deformed' Yangian symmetry
one may hope that perhaps the Yangian algebra implies some constraints on higher-point amplitudes.
In particular it is not impossible that such constraints could help to fix the two-loop
six-gluon MHV amplitude, which at the moment is known in terms of rather complicated parametric
integrals only \cite{Drummond:2007bm,Drummond:2008aq,Bern:2008ap,Anastasiou:2009kn}.

On-shell scattering amplitudes and the spectrum of
anomalous dimensions of gauge-invariant operators in $\mathcal{N}=4$ SYM
exhibit hidden symmetries.
Given these findings it seems worthwhile to look for new symmetries of other objects
in $\mathcal{N}=4$ SYM as well. Indeed, a Yangian symmetry could have easily been overlooked
in previous studies since the higher-order Yangian generators are intrinsically non-local.
It would be wonderful if e.g.~three- and four-point functions
of gauge-invariant composite operators were governed by a similar Yangian symmetry.

\subsection*{Acknowledgements}
We thank Luc Frappat, Axel Kleinschmidt, Thomas Quella, Eric Ragoucy, Emery Sokatchev, Fabian Spill, 
and particularly Niklas Beisert and Edward Witten for important discussions.
This research was supported in part by the French Agence Nationale de la
Recherche under grant ANR-06-BLAN-0142 and by the Volkswagen Foundation.

\section*{Appendix: Formulae for both superconformal algebras}
We
begin by listing the commutation relations of the algebra $u(2,2|4)$. The Lorentz generators
$\mathbb{M}_{\a \b}$, $\overline{\mathbb{M}}_{\da \db}$ and the $su(4)$ generators
$\mathbb{R}^{A}{}_{B}$ act canonically on the remaining generators carrying Lorentz or $su(4)$
indices. The dilatation $\mathbb{D}$ and hypercharge $\mathbb{B}$ act via
\be
[\mathbb{D},\mathbb{J}] = {\rm dim}(\mathbb{J})\, \mathbb{J}, \qquad [\mathbb{B},\mathbb{J}] = {\rm
hyp}(\mathbb{J})\, \mathbb{J}.
\ee
The non-zero dimensions and hypercharges of the various generators are
\begin{align} \notag
& {\rm dim}(\mathbb{P})=1, \qqquad {\rm dim}(\mathbb{Q}) = {\rm dim}(\overline{\mathbb{Q}}) =
\tfrac{1}{2},\qquad {\rm dim}(\mathbb{S}) = {\rm dim}(\overline{\mathbb{S}}) = -\tfrac{1}{2}
\\
&{\rm dim}(\mathbb{K})=-1,\qquad {\rm hyp}(\mathbb{Q}) = {\rm hyp}(\overline{\mathbb{S}}) =
\tfrac{1}{2}, \qquad~ {\rm hyp}(\overline{\mathbb{Q}}) = {\rm hyp}(\mathbb{S}) = - \tfrac{1}{2}.
\end{align}
The remaining non-trivial commutation relations are,
\begin{align} \notag
& \{\mathbb{Q}_{\a A},\overline{\mathbb{Q}}_{\da}^B\}  =  \delta_A^B \mathbb{P}_{\a \da},
   \qquad \{\mathbb{S}_{\a}^A,\overline{\mathbb{S}}_{\da B} \} = \delta_B^A \mathbb{K}_{\a \da},
\\ \notag
& {}[\mathbb{P}_{\a \da},\mathbb{S}^{\b A}] = \delta_{\a}^{\b} \overline{\mathbb{Q}}_{\da}^A,
 \qqquad [\mathbb{K}_{\a \da},\mathbb{Q}^{\b}_{A}] = \delta_{\a}^{\b}
   \overline{\mathbb{S}}_{\da A},
\\ \notag
& {}[\mathbb{P}_{\a \da},\overline{\mathbb{S}}^{\db}_{A}]  =  \delta^{\db}_{\da} \mathbb{Q}_{\a A},
\qqquad [\mathbb{K}_{\a \da}, \overline{\mathbb{Q}}^{\db A}]  =  \delta_{\da}^{\db} \mathbb{S}_{\a}^{A},
\\ \notag
& [\mathbb{K}_{\a \da},\mathbb{P}^{\b \db}] = \delta_\a^\b \delta_\da^\db \mathbb{D} +
\mathbb{M}_{\a}{}^{\b}
 \delta_\da^\db + \overline{\mathbb{M}}_{\da}{}^{\db} \delta_\a^\b,
\\ \notag
& \{\mathbb{Q}^{\a}_{A},\mathbb{S}_\b^B\} =  \mathbb{M}^{\a}{}_{\b}
\delta_A^B + \delta^{\a}_{\b} \mathbb{R}^{B}{}_{A} + \tfrac{1}{2}\delta^{\a}_{\b} \delta_A^B (\mathbb{D}+\mathbb{C}),
\\ \label{comm-rel}
& \{\overline{\mathbb{Q}}^{\da A},\overline{\mathbb{S}}_{\db B}\} = \overline{\mathbb{M}}^{\da}{}_{\db} \delta_B^A  - \delta^{\da}_{\db} \mathbb{R}^{A}{}_{B} + \tfrac{1}{2} \delta^{\da}_{\db}\delta_B^A
(\mathbb{D}-\mathbb{C}).
\end{align}
Note that in writing the algebra relations we are obliged to choose the $su(4)$ chirality of the odd generators. The relations above are valid directly for the dual superconformal generators. For the conventional realisation of the algebra, one should simply swap all $su(4)$ chiralities appearing in the commutation relations.
We now give the generators in both the conventional and dual representations of the superconformal
algebra. We will use the following shorthand notation:
\begin{align}\label{shortderiv}
\partial_{i \alpha \dot{\alpha}} = \frac{\partial}{\partial
x_i^{\alpha \dot{\alpha}}}, \qquad \partial_{i \alpha A} = \frac{\partial}{\partial \theta_i^{\alpha
A}}, \qquad \partial_{i \alpha} = \frac{\partial}{\partial \lambda_i^{\alpha}}\,, \qquad
\partial_{i \dot{\alpha}} = \frac{\partial}{\partial
    \tilde{\lambda}_i^{\dot{\alpha}}}\,, \qquad
\partial_{i A} = \frac{\partial}{\partial \eta_i^A}\,.
\end{align}
We first give the generators of the conventional superconformal symmetry, using lower case
characters to distinguish these generators from the dual superconformal generators which follow
afterwards.
\begin{align}
& p^{\dot{\alpha}\alpha }  =  \sum_i \tilde{\lambda}_i^{\dot{\alpha}}\lambda_i^{\alpha} \,, & &
k_{\alpha \dot{\alpha}} = \sum_i \partial_{i \alpha} \partial_{i \dot{\alpha}} \,,\notag\\
&\overline{m}_{\dot{\alpha} \dot{\beta}} = \sum_i \tilde{\lambda}_{i (\dot{\alpha}} \partial_{i
\dot{\beta} )}, & & m_{\alpha \beta} = \sum_i \lambda_{i (\alpha} \partial_{i \beta )}
\,,\notag\\
& d =  \sum_i [\tfrac{1}{2}\lambda_i^{\alpha} \partial_{i \alpha} +\tfrac{1}{2}
\tilde{\lambda}_i^{\dot{\alpha}} \partial_{i
    \dot{\alpha}} +1], & & r^{A}{}_{B} = \sum_i [-\eta_i^A \partial_{i B} + \tfrac{1}{4}\delta^A_B \eta_i^C \partial_{i C}]\,,\notag\\
&q^{\alpha A} =  \sum_i \lambda_i^{\alpha} \eta_i^A \,, &&   \bar{q}^{\dot\alpha}_A
= \sum_i \tilde\lambda_i^\da \partial_{i A} \,, \notag\\
& s_{\alpha A} =  \sum_i \partial_{i \alpha} \partial_{i A}, & &
\bar{s}_{\dot\alpha}^A = \sum_i \eta_i^A \partial_{i \dot\alpha}\,,\notag\\
&c = \sum_i [1 + \tfrac{1}{2} \lambda_i^{\a} \partial_{i \a} - \tfrac{1}{2} \tilde\lambda^{\da}_i \partial_{i \da} - \tfrac{1}{2} \eta^A_i \partial_{iA} ]\,.
\end{align}
We can construct the generators of dual superconformal transformations by starting with the standard
chiral representation and extending the generators so that they commute with the constraints,
\be
(x_i-x_{i+1})_{\a \dot\alpha}  - \lambda_{i\, \a}\, \tilde{\lambda}_{i\, \dot\alpha} = 0\,, \qquad (\theta_i - \theta_{i+1})_\alpha^A - \lambda_{i \alpha} \eta_i^A = 0\,.
\ee
By construction they preserve the surface defined by these constraints, which is where the amplitude
has support. The generators are
\begin{align}
P_{\alpha \dot{\alpha}}&= \sum_i \partial_{i \alpha \dot{\alpha}}\,, \qquad Q_{\alpha A} = \sum_i \partial_{i \alpha A}\,, \qquad
\overline{Q}_{\dot{\alpha}}^A = \sum_i [\theta_i^{\alpha A}
  \partial_{i \alpha \dot{\alpha}} + \eta_i^A \partial_{i \dot{\alpha}}],\label{barQfss} \\
M_{\alpha \beta} &= \sum_i[x_{i ( \alpha}{}^{\dot{\alpha}}
  \partial_{i \beta ) \dot{\alpha}} + \theta_{i (\alpha}^A \partial_{i
  \beta) A} + \lambda_{i (\alpha} \partial_{i \beta)}]\,, \qquad
\overline{M}_{\dot{\alpha} \dot{\beta}} = \sum_i [x_{i
    (\dot{\alpha}}{}^{\alpha} \partial_{i \dot{\beta} ) \alpha} +
  \tilde{\lambda}_{i(\dot{\alpha}} \partial_{i \dot{\beta})}]\,,\\
R^{A}{}_{B} &= \sum_i [\theta_i^{\alpha A} \partial_{i \alpha B} +
  \eta_i^A \partial_{i B} - \tfrac{1}{4} \delta^A_B \theta_i^{\alpha
    C} \partial_{i \alpha C} - \tfrac{1}{4}\delta^A_B \eta_i^C \partial_{i C}
]\,,\\
\label{DD} D &= \sum_i [-x_i^{\dot{\alpha}\alpha}\partial_{i \alpha \dot{\alpha}} -
  \tfrac{1}{2} \theta_i^{\alpha A} \partial_{i \alpha A} -
  \tfrac{1}{2} \lambda_i^{\alpha} \partial_{i \alpha} -\tfrac{1}{2}
  \tilde{\lambda}_i^{\dot{\alpha}} \partial_{i \dot{\alpha}}]\,,\\
  \label{CC}
C &=  \sum_i [-\tfrac{1}{2}\lambda_i^{\alpha} \partial_{i \alpha} +
  \tfrac{1}{2}\tilde{\lambda}_i^{\dot{\alpha}} \partial_{i \dot{\alpha}} + \tfrac{1}{2}\eta_i^A
  \partial_{i A}]\,, \\
S_{\alpha}^A &= \sum_i [-\theta_{i \alpha}^{B} \theta_i^{\beta A}
  \partial_{i \beta B} + x_{i \alpha}{}^{\dot{\beta}} \theta_i^{\beta
    A} \partial_{i \beta \dot{\beta}} + \lambda_{i \alpha}
  \theta_{i}^{\gamma A} \partial_{i \gamma} + x_{i+1\,
    \alpha}{}^{\dot{\beta}} \eta_i^A \partial_{i \dot{\beta}} -
  \theta_{i+1\, \alpha}^B \eta_i^A \partial_{i B}]\,,\\
\overline{S}_{\dot{\alpha} A} &= \sum_i [x_{i \dot{\alpha}}{}^{\beta}
  \partial_{i \beta A} + \tilde{\lambda}_{i \dot{\alpha}}
  \partial_{iA}]\,,\label{fssbarS}\\
\label{KK}
K_{\alpha \dot{\alpha}} &= \sum_i [x_{i \alpha}{}^{\dot{\beta}} x_{i
    \dot{\alpha}}{}^{\beta} \partial_{i \beta \dot{\beta}} + x_{i
    \dot{\alpha}}{}^{\beta} \theta_{i \alpha}^B \partial_{i \beta B} +
  x_{i \dot{\alpha}}{}^{\beta} \lambda_{i \alpha} \partial_{i \beta}
  + x_{i+1 \,\alpha}{}^{\dot{\beta}} \tilde{\lambda}_{i \dot{\alpha}}
  \partial_{i \dot{\beta}} + \tilde{\lambda}_{i \dot{\alpha}} \theta_{i+1\,
    \alpha}^B \partial_{i B}]\,.
\end{align}
Here the summations go from $1$ to $n$ if we identify the dual superspace points $(x_{1},\theta_{1})=(x_{n+1},\theta_{n+1})$.
However, as was discussed in the text, if we want to act on an amplitdue as a distribution,
we need to consider $(x_{1},\theta_{1})\neq(x_{n+1},\theta_{n+1})$ instead. In this case
the terms involving derivatives $\partial / \partial x_{i}$ and  $\partial / \partial \theta_{i}$ are
summed from $1$ to $n+1$.

Note that if we restrict the dual generators $\bar{Q},\bar{S}$ to the on-shell superspace they
become identical to the conventional generators $\bar s, \bar q$.

\subsection*{The fundamental representation of $u(2,2|4)$}

First let us give the fundamental representation of $u(2,2|4)$. This representation is given by $(4|4)\times (4|4)$ matrices. We will use the notation $E^{\underline A}{}_{\underline B}$ to refer to a matrix which is zero everywhere except for the entry $1$ in row $\underline A$ and column $\underline B$.  We will denote the fundamental representation of of the generator $J_a$ by $M[J_a]$. Let us arrange all generators except $d,c,b$ into a matrix. Then the non-zero entries of these generators are given by
\begin{align}
M \left[ \left (
\begin{matrix}
m^{\alpha}{}_{\beta} & p^{\alpha}{}_{\dot\beta}& q^{\alpha B} \\
k^{\dot\alpha}{}_{\beta}& \overline{m}^{\dot\alpha}{}_{\dot\beta} & \bar{s}^{\dot\alpha B} \\
s_{A \beta}& \bar{q}{}_{A \dot\beta}& r_{A}{}^{B}
\end{matrix}
\right )\right] =
\left(
\begin{matrix}
E^{\alpha}{}_{\beta} - \tfrac{1}{2}\delta^\alpha_\beta \mathbb{I} &
E^{\alpha}{}_{\dot\beta}& E^{\alpha B} \\
E^{\dot\alpha}{}_{\beta}& E^{\dot\alpha}{}_{\dot\beta} - \tfrac{1}{2} \delta^{\dot\alpha}_{\dot\beta} \mathbb{I} & E^{\dot\alpha B} \\
E_{A \beta}& E{}_{A \dot\beta}& E_{A}{}^{B} - \tfrac{1}{4} \delta_A^B \mathbb{I}
\end{matrix}
\right).
\end{align}
This equation is to be read as
\be
M[m^{\alpha}{}_{\beta}]=\left (
\begin{matrix}
E^{\alpha}{}_{\beta} - \tfrac{1}{2}\mathbb{I}\delta^\alpha_\beta &0&0 \\
0& 0 & 0 \\
0& 0& 0
\end{matrix}
\right )
\ee
and so on.
The remaining generators are given by
\begin{align}
M[d]=
\left(
\begin{matrix}
\tfrac{1}{2}\mathbb{I} &0&0 \\
0& -\tfrac{1}{2} \mathbb{I} &0  \\
0&0&0
\end{matrix}
\right),
\hspace{10pt}
M[c]=
\left(
\begin{matrix}
\tfrac{1}{2}\mathbb{I} &0&0 \\
0& \tfrac{1}{2} \mathbb{I} &0  \\
0&0& \tfrac{1}{2} \mathbb{I}
\end{matrix}
\right),
\hspace{10pt}
M[b]=
\left(
\begin{matrix}
0&0&0 \\
0& 0 &0  \\
0&0&- \tfrac{1}{2} \mathbb{I}
\end{matrix}
\right).
\end{align}
One can easily check that these matrices obey the algebra relations.

The metric is defined by
\be
g_{ab}= g(J_a,J_b) = {\rm str}(M[J_a] M[J_b]).
\ee
It satisfies
\be
g_{ba} = (-1)^{|a|} g_{ab}, \hspace{20pt} g_{ab}=0 \text{ if } |a| \neq |b|.
\ee
The non-zero components of the metric are
\begin{align}
&g(p^{\alpha}{}_{\dot\beta},k^{\dot\gamma}{}_{\delta}) = g(k^{\dot\gamma}{}_{\delta},p^{\alpha}{}_{\dot\beta})=\delta^\alpha_\delta \delta^{\dot\gamma}_{\dot \beta}, \quad \,\,\,\, \, g(r_{A}{}^{B},r_{C}{}^{D}) = -\delta_A^D \delta_C^B + \tfrac{1}{4}\delta_A^B \delta_C^D,\notag\\
&g(m^{\alpha}{}_{\beta},m^{\gamma}{}_{\delta}) = \delta^{\alpha}_{\delta} \delta^{\gamma}_{\beta} - \tfrac{1}{2} \delta^{\alpha}_{\beta} \delta^{\gamma}_{\delta},  \qqquad  \, \,  g(\overline{m}^{\dot\alpha}{}_{\dot\beta},\overline{m}^{\dot\gamma}{}_{\dot\delta}) = \delta^{\dot\alpha}_{\dot\delta} \delta^{\dot\gamma}_{\dot\beta} - \tfrac{1}{2} \delta^{\dot\alpha}_{\dot\beta} \delta^{\dot\gamma}_{\dot\delta}, \notag \\
&g(d,d) = 1,   \qqquad \qqquad \qqquad \qquad g(c,b) = g(b,c)  = 1, \notag\\
&g(q^{\alpha B}, s_{C \delta}) = - g(s_{C \delta},q^{\alpha B})  = \delta^B_C \delta^\alpha_\delta  \quad g(\bar{s}^{\dot\alpha B},\bar{q}_{C\dot\delta}) = - g(\bar{q}_{C\dot\delta},\bar{s}^{\dot\alpha B}) = \delta_C^B \delta^{\dot\alpha}_{\dot\delta}.
\end{align}
In fact it is convenient to use a different basis of generators. We will define the generators
\be
y^{\alpha}{}_{\beta} = m^{\alpha}{}_{\beta} + \tfrac{1}{2}\delta^\alpha_\beta(d+c+b),\qquad \overline{y}^{\dot\alpha}{}_{\dot\beta} = \overline{m}^{\dot\alpha}{}_{\dot\beta} + \tfrac{1}{2} \delta^{\dot\alpha}_{\dot\beta}(c+b-d), \qquad w_{A}{}^{B} = r_{A}{}^{B} -\tfrac{1}{2} \delta^B_A b
\ee
Then the fundamental representation is easy to write. We organise the generators into a matrix
\be
M[J^{\underline A}{}_{\underline B}] =
M\left[
\left(
\begin{matrix}
y^{\alpha}{}_{\beta}&p^{\alpha}{}_{\dot\beta}&q^{\alpha B} \\
k^{\dot\alpha}{}_{\beta}& \overline{y}^{\dot\alpha}{}_{\dot\beta} &\bar{s}^{\dot\alpha B}  \\
s_{A \beta}& \bar{q}_{A \dot\beta} & w_{A}{}^{B}
\end{matrix}
\right)
\right]
= E^{\underline A}{}_{\underline B}
\ee
In this basis the non-zero components of the metric are
\begin{align}
&g(p^{\alpha}{}_{\dot\beta},k^{\dot\gamma}{}_{\delta}) = g(k^{\dot\gamma}{}_{\delta},p^{\alpha}{}_{\dot\beta})=\delta^\alpha_\delta \delta^{\dot\gamma}_{\dot \beta}, \qquad  g(w_{A}{}^{B},w_{C}{}^{D}) = -\delta_A^D \delta_C^B,\notag\\
&g(y^{\alpha}{}_{\beta},y^{\gamma}{}_{\delta}) = \delta^{\alpha}_{\delta} \delta^{\gamma}_{\beta},  \qqquad  \qqquad  \qquad g(\overline{y}^{\dot\alpha}{}_{\dot\beta},\overline{y}^{\dot\gamma}{}_{\dot\delta}) = \delta^{\dot\alpha}_{\dot\delta} \delta^{\dot\gamma}_{\dot\beta}, \notag \\
&g(q^{\alpha B}, s_{C \delta}) = - g(s_{C \delta},q^{\alpha B})  = \delta^B_C \delta^\alpha_\delta  \quad \,\, g(\bar{s}^{\dot\alpha B},\bar{q}_{C\dot\delta}) = - g(\bar{q}_{C\dot\delta},\bar{s}^{\dot\alpha B}) = \delta_C^B \delta^{\dot\alpha}_{\dot\delta}.
\end{align}

We will denote the inverse metric by $g^{ab}  = g^{-1}(J_a,J_b)$. It satisfies
\be
g_{ab}g^{bc} = \delta_a^c = g^{cb}g_{ba}, \hspace{30pt} g^{ba} = (-1)^{|a|} g^{ab}.
\ee
The non-zero components of the inverse metric are best written in the matrix basis,
\begin{align}
&g^{-1}(p^{\alpha}{}_{\dot\beta},k^{\dot\gamma}{}_{\delta}) = g^{-1}(k^{\dot\gamma}{}_{\delta},p^{\alpha}{}_{\dot\beta}) = \delta_\alpha^\delta \delta^{\dot\beta}_{\dot\gamma},  \qquad \,\,\,\,\, g^{-1}(w_{A}{}^{B},w_{C}{}^{D}) = -\delta^A_D \delta^C_B,\notag\\
&g^{-1}(y^{\alpha}{}_{\beta},y^{\gamma}{}_{\delta}) = \delta_{\alpha}^{\delta} \delta_{\gamma}^{\beta},  \qqquad  \qqquad  \qqquad \,\,\,\,\,  g^{-1}(\overline{y}^{\dot\alpha}{}_{\dot\beta},\overline{y}^{\dot\gamma}{}_{\dot\delta}) = \delta_{\dot\alpha}^{\dot\delta} \delta_{\dot\gamma}^{\dot\beta}, \notag \\
&g^{-1}(q^{\alpha B}, s_{C \delta}) = - g^{-1}(s_{C \delta},q^{\alpha B})  = - \delta_B^C \delta_\alpha^\delta  \quad \,\,\, g^{-1}(\bar{s}^{\dot\alpha B},\bar{q}_{C\dot\delta}) = - g^{-1}(\bar{q}_{C\dot\delta},\bar{s}^{\dot\alpha B}) = -\delta^C_B \delta_{\dot\alpha}^{\dot\delta}.
\end{align}

We will raise and lower indices with the left index of $g_{ab}$ or $g^{ab}$. Thus, we will define the lowered structure constants,
\be
f_{abc} = f_{ab}{}^{d}g_{dc}.
\ee
The metric is invariant, i.e.
\be
g([J_a,J_b],J_c) = g(J_a,[J_b,J_c]).
\ee
Therefore we have
\be
f_{abc} = f_{bc}{}^{d}g_{ad} = (-1)^{|a|} f_{bca}.
\ee
Using these properties we find
\begin{align}
f_{a}{}^{cb} f_{bc}{}^{d} &= (-1)^{|d|}f_{ac'}{}^{b} f^{d}{}_{bc} g^{c'c} \notag \\
&= (-1)^{|d| + |c|} f_{ac}{}^{b}f^{d}{}_{b}{}^{c} 
\label{ffvanishing}
\end{align}
This expression vanishes for certain super-algebras, including $psu(n|n)$, due to the vanishing of the Killing form,
\be
K_{ab}={\rm str}[{\rm ad}(J_a) {\rm ad}(J_{b})] = (-1)^{|c|} f_{ac}{}^{d} f_{bd}{}^{c}.
\label{killingform}
\ee
Let us now check that the formula
\be
J^{(1)}_a = f_{a}{}^{cb} \sum_{i<j} J_{ib} J_{jc}
\label{J1formula}
\ee
agrees with our explicit expression for $q^{(1)}$.
We find from (\ref{J1formula})
\begin{align}
q^{(1) \alpha A} &= f_{q^{\alpha A}}{}^{cb} \sum_{i<j} J_{ib} J_{jc} \notag \\
&= f_{q^{\alpha A} c'}{}^{b} g^{c'c} \sum_{i<j} J_{ib} J_{jc} 
\label{q1calc1}
\end{align}
The non-zero structure constants contribute the following terms,
\begin{align}
&f_{q^{\alpha A} \bar{q}_{B \dot\alpha}}{}^{p^{\beta}{}_{\dot\beta}} g^{-1}(\bar{q}_{B \dot\alpha} , \bar{s}^{\dot \gamma C}) \sum_{i<j} p_{i}^{\beta}{}_{\dot\beta} \bar{s}_j^{\dot \gamma C} 
+ f_{q^{\alpha A} k^{\dot\alpha}{}_{\beta}}{}^{\bar{s}^{\dot\gamma C}} g^{-1}(k^{\dot\alpha}{}_{\beta} , p^{\gamma}{}_{\dot\delta}) \sum_{i<j} \bar{s}_i^{\dot\gamma C} p_j^{\gamma}{}_{\dot\delta} \notag \\
+& f_{q^{\alpha A} s_{B \beta}}{}^{y^{\gamma}{}_{\delta}} g^{-1}(s_{B\beta} , q^{\rho C}) \sum_{i<j} y_i^{\gamma}{}_{\delta} q_{j}^{\rho C} + f_{q^{\alpha A} y^{\gamma}{}_{\delta}}{}^{q^{\epsilon E}} g^{-1}(y^{\gamma}{}_{\delta} , y^{\rho}{}_{\lambda}) \sum_{i<j} q_i^{\epsilon E} y_j^{\rho}{}_{\lambda} \notag \\
+&f_{q^{\alpha a} s_{B\beta}}{}^{w_{D}{}^{C}} g^{-1}(s_{B\beta} , q^{\rho E}) \sum_{i<j} w_{iD}{}^{C} q_j^{\rho E} + f_{q^{\alpha A} w_{B}{}^{C}}{}^{q^{\beta D}} g^{-1}(w_{B}{}^{C},w_{F}{}^{E}) \sum_{i<j} q_i^{\beta D} w_{jF}{}^{E} \notag \\
=& \sum_{i<j} \Bigl[p_i^{\alpha}{}_{\dot\alpha} \bar{s}_j^{\dot\alpha A} + y_i^{\alpha}{}_{\beta} q_j^{\beta A} + w_{iB}{}^{A} q_j^{\alpha B} - (j,i)\Bigr] \notag \\
=& \sum_{i<j} \Bigl[p_i^{\alpha}{}_{\dot\alpha} \bar{s}_j^{\dot\alpha A} + (m_i^{\alpha}{}_{\beta} + \tfrac{1}{2} \delta^{\alpha}_{\beta} (d_i+c_i+b_i))q_j^{\beta A} + (r_{iB}{}^{A} - \tfrac{1}{2} \delta^A_B b_i) q_j^{\alpha B} - (j,i) \Bigr]
\label{q1calc2}
\end{align}
The terms involving the hypercharge density $b_j$ cancel out and we find agreement with formula (\ref{q1level1final}).

\end{document}